\documentclass[preprint,aps]{revtex4}
\usepackage{graphics}
\usepackage{comment}
\usepackage{bm} 
\usepackage{amssymb,amsmath,
amsthm,epsfig,subfigure}

 \makeatletter  \newcommand{\Rmnum}[1]{\expandafter\@slowromancap\romannumeral #1@}

\usepackage[belowskip=-30pt]{caption}
\makeatother

\begin{document}

\title{Information length as a useful index to understand variability in the global circulation}
\author{Eun-jin Kim$^{1,2}$, James Heseltine$^{1,2}$, Hanli Liu$^3$}
\affiliation{
$^1$School of Mathematics and Statistics,
University of Sheffield, Sheffield, S3 7RH, UK\\
$^{2}$ Fluid and Complex Systems Research Centre, Coventry University, Coventry CV1 2TT, UK\\
$^3$ High Altitude Observatory, National Centre for Atmospheric Research, Boulder, CO80303-3000, USA
}
\vspace{1cm}
\begin{abstract}
With improved measurement and modelling technology, variability has emerged as an essential feature in non-equilibrium processes.
While traditionally, mean values and variance have been heavily used, they are not appropriate in describing extreme events where a significant deviation from mean values often occurs. Furthermore, stationary Probability Density Functions (PDFs) miss crucial information about the dynamics associated with variability. It is thus critical to go beyond a traditional approach and deal with time-dependent PDFs. 
Here, we consider atmospheric data from the Whole Atmosphere Community Climate Model (WACCM) and calculate time-dependent PDFs and the information length from these PDFs, which is the total number of statistically different states that a system passes through in time. Time-dependent PDFs are shown to be non-Gaussian in general, and the information length calculated from these PDFs shed us a new 
perspective of understanding variabilities, correlation among different variables and regions. 
Specifically, we calculate time-dependent PDFs and information length and show that the information length tends to increase with the altitude albeit in a complex form. This tendency is more robust for flows/shears than temperature. Also, much similarity among flows and shears in the information length is found in comparison with the temperature. This means a stronger correlation among flows/shears because of a strong coupling through gravity waves in this particular WACCM model. We also find the increase of the information length with the latitude and interesting hemispheric asymmetry for flows/shears/temperature, a stronger anti-correlation (correlation) between flows/shears and temperature at a higher (low) latitude.
These results also suggest the importance of high latitude/altitude in the information budge in the Earth's atmosphere, the spatial gradient of the information as a useful proxy for the transport of physical quantities.

\end{abstract}

\maketitle

\section{Introduction}
With improved measurement and modelling technology, variability has emerged as an essential feature in non-equilibrium processes.
Closely linked to unpredictability, variability plays a crucial role in various unexpected or undesirable events such as fusion plasma eruption, extreme weather conditions, stock market crash, etc \cite{nature,KIM02,KIM03,KIM06,KIM08,KIM13,
SrinYoun2011,SayaShowDowl2008,tsuchiya15,tang88,Jensen98,Pruesser12,Longo11,
FLYNN2014,FLYNN2015,Ovidiu,Shahrezaei,Thomas,Biswas,Elgart}. 
Specifically,
anomalous (much larger than mean values) transport associated with large
fluctuations in fusion plasmas can degrade the confinement, potentially even
terminating fusion operation \cite{KIM03}. Tornadoes are rare, large amplitude
events, but can cause very substantial damage when they do occur. Furthermore,
gene expression and protein productions, which used to be thought of as smooth
processes, have also been observed to occur in bursts
(e.g.\ \cite{Ovidiu,Shahrezaei,Thomas,Biswas,Elgart}. 
%Such rare events of large
%amplitude (called intermittency) can dominate the entire transport even if they
occur infrequently \cite{KIM08,KIM09}.

How to quantify variability mathematically however does not seem to be well established. For unpredictable events, we use a Probability Density Function (PDF) to describe the likelihood of a certain event to take place. A simplest and popular example is Gaussian PDF which has the nice property of symmetry and uniquely being defined by only two parameters -- the mean value $\mu$ for the peak position and standard deviation $\sigma$ for the width of a PDF. Note that the variance is the square of standard deviation.
As a broad PDF has a wide range of values for a finite probability, suggesting less predictability,  variability can mean a large variance. On the other hand, the temporal change in mean value is also used as a measure of variability. How do then we treat the case where  
variance increases or decreases in time? Furthermore, for a non-Gaussian PDF, we also need to consider the change in other characteristics like symmetry, skewness or kurtosis, and all other higher moment. 
This is especially important for extreme events noted above since the assumption of
small fluctuations with short correlation time for the Gaussian PDF badly fails, with a very limited utility of mean value and variance.
This brings us the importance of considering the entire PDF and their time evolution in
defining variability. \\
%This is especially important for extreme events where a significant deviation from mean values often occurs.\\

In our previous work, we showed that time-dependent PDFs provide a key insight that is completely missing in any studies using only mean values, variance or stationary PDFs. Specifically, we quantify the similarity and disparity between PDFs by assigning the metric between the two such that the distance between two PDFs increases with the disparity between them \cite{WOOTTERS81,fisher}. For Gaussian PDFs, a statistically different state is attained when the physical distance exceeds the resolution set by the uncertainty (PDF width). We extended this concept to time-dependent problems where a PDF changes continuously in time and introduced the information length ${\cal L}$ to quantify the number of statistically different states that a system passes through in time to reach time $t$ starting from an initial PDF at time $0$ \cite{KH16,KH17,Entropy,paper6,NK14,NK15,HK16,KIM16}.
One of the merits of ${\cal L}$ is that it is invariant under the (time-independent) change of variables and thus can be directly compared between different variables unlike physical variables which have different units. For instance, it can be used to quantify the correlation between different variables \cite{Heseltine19}.

Rigorously, ${\cal L}$ can be shown to be related to the sum of the infinitesimal relative entropy along the trajectory of the system \cite{KIM16,KIM18}. 
It is however instructive to consider defining i) a dynamical time scale $\tau(t)$ as the rate of information change and then ii) by measuring
the clock time $t$ by $\tau$.
For example, for a time dependent PDF $p(x,t)$, $\tau$ is calculated as
\begin{eqnarray}
\frac{1}{\tau^2} & = &\int dx \frac {1} {p(x,t)}
 \left [\frac {\partial p(x,t)} {\partial t} \right]^2.
 \label{eq1}
 \end{eqnarray}
From Eq.\ (\ref{eq1}), we can see that the dimension of $\tau=\tau(t)$ is time and serves as a dynamical time unit for information change.
${\cal L}(t)$ is the total information change between time $0$ and~$t$:
\begin{eqnarray}
{\cal{L}} (t) &=& \int_0^{t} \frac{dt_1}{\tau(t_1)}
 = \int_0^{t} dt_1 \sqrt{\int dx \frac {1} {p(x, t_1)}
 \left [\frac {\partial p(x,t_1)} {\partial t_1} \right]^2}.
\label{eq2}
\end{eqnarray}
The integral in Eq.\ (\ref{eq2}) is necessary since $\tau(t)$ in Eq.\ (\ref{eq1}) depends on time in general.
To understand this, we can consider an oscillator with the characteristic time scale $
\tau$ given by its period $\tau = 2$ secs. Then, within the clock time 10 secs, the number of 5 oscillation would correspond to the information length 5. In the case where the period $\tau$ varies with time, what
is required is the integral of $1/\tau$ over the time. 

As a measure of the information change, 
${\cal L}_\infty$ was shown to map out an attractor structure. In particular, 
in the case of a stable equilibrium, the effect of different deterministic forces was demonstrated by the scaling of ${\cal L}_\infty$ 
against the peak position of a narrow initial PDF, 
the minimum value of ${\cal L}_\infty$ occurring at the equilibrium point.
Furthermore, ${\cal L}_{\infty}$ varies smoothly with the initial conditions (e.g. the distance of an initial PDF from the attractor point).
In a sharp contrast, in the case of a chaotic attractor, ${\cal L}_{\infty}$ varies abruptly with the peak position of a narrow initial PDF;
this sensitive dependence on initial conditions is reminiscent of a Lyapunov exponent. 
That is, ${\cal L}$ provides a new way of understanding dynamical systems.
Finally, the information length can also be applied to any data such as music (e.g.\ see \cite{NK15}) where the information flow in
different classical musics (e.g.\ see \cite{NK15}) were calculated. 

In this work, we apply this to the Whole Atmosphere Community Climate Model (WACCM) and show that 
the information length ${\cal L}(t)$ as a useful index to measure `dynamic variability' and `correlation'.
The remainder of this paper is organized as follows. Section II provides
the analysis of WACCM data. Discussions and Conclusion are provided in Section III.

\section{Earth Atmosphere from the WACCM}

The Whole Atmosphere Community Climate Model (WACCM) is a global circulation model which has been developed for the last few decades through an inter-divisional collaboration at the National Center for Atmospheric Research (NCAR). Specifically, it integrates the upper atmospheric modeling of High Altitude Observatory, the middle atmosphere modeling of Atmospheric Chemistry Observations \& Modeling, and the tropospheric modeling of Climate \& Global Dynamics, using the NCAR Community Earth System Model as a common numerical framework. The WACCM is a comprehensive numerical model, spanning the range of altitude from the Earth's surface to the thermosphere. The particular data that we analyse are described in details in \cite{Liu1,Liu2}.

We are interested in the information budget in 7 layers in the atmosphere covering thermosphere, mesopause, mesosphere, stratopause, stratosphere, tropopause, troposphere from the top to the bottom of the atmosphere.
We consider three different cases of data sampling for PDFs:
\begin{itemize}
\item All longitude and latitude data to understand the global information budge at 7 layers.
\item All longitude data to understand the information budget across latitude at 7 layers.
\item All longitude and latitude data to understand the global information budget at all altitude. 
\end{itemize}
According to the sampling in each case above,
we calculate the time-dependent PDFs for the six variables, 
\begin{itemize}
\item Temperature $T$.
\item Zonal flow $U$.
\item Meridional flow $V$.
\item Zonal (vertical) shear $\frac{\partial U}{\partial z}$. 
\item Meridional (vertical) shear $\frac{\partial V}{\partial z}$.
\item Total (vertial) shear $\sqrt{\left( \frac{\partial U}{\partial z} \right)^2
+ \left(\frac{\partial V}{\partial z} \right)^2}$. 
\end{itemize}

\subsection{Information budget at the 7 layers}
\begin{figure}
\includegraphics[angle=0,width=5.4cm,height=7cm]{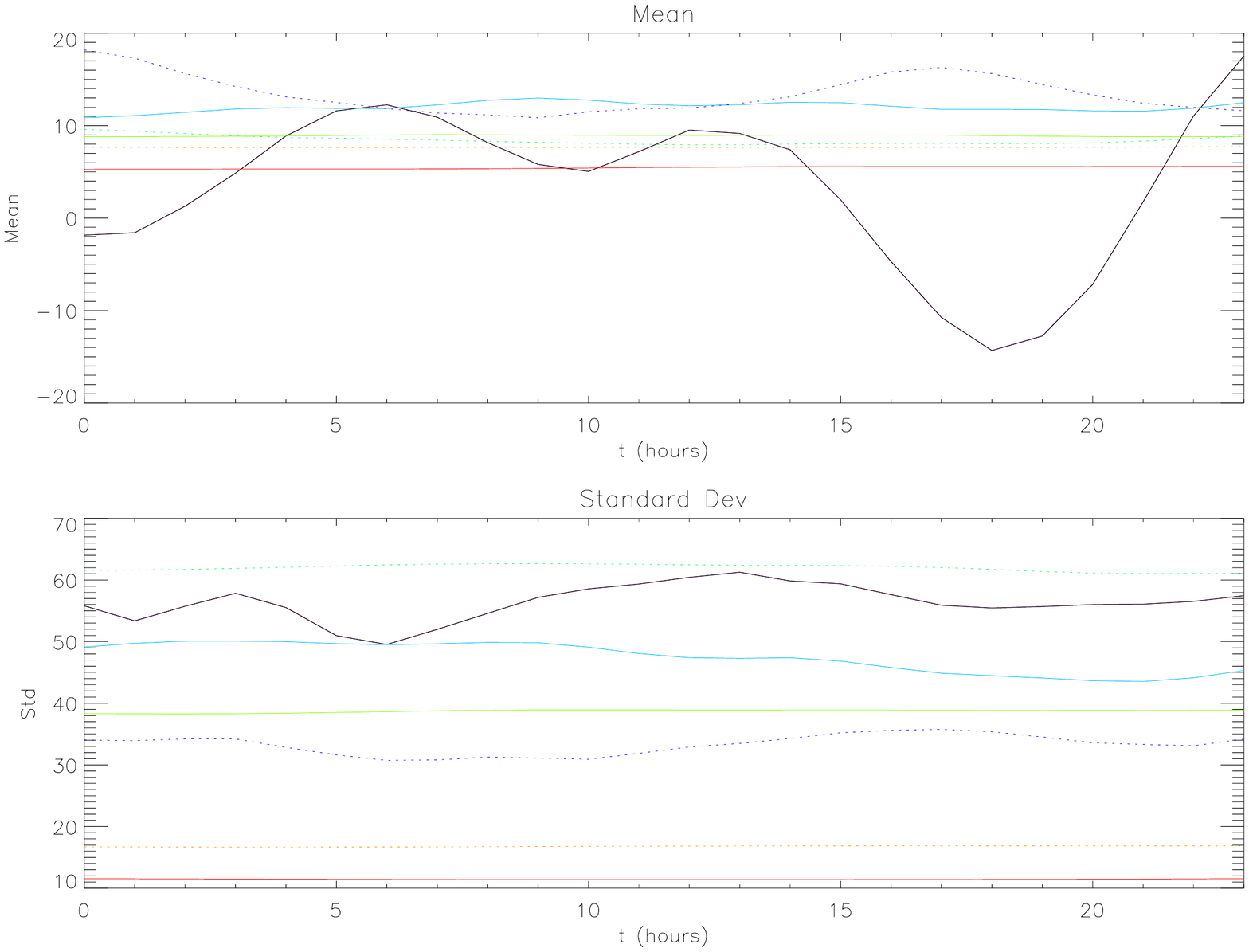}
\includegraphics[angle=0,width=5.4cm,height=7cm]{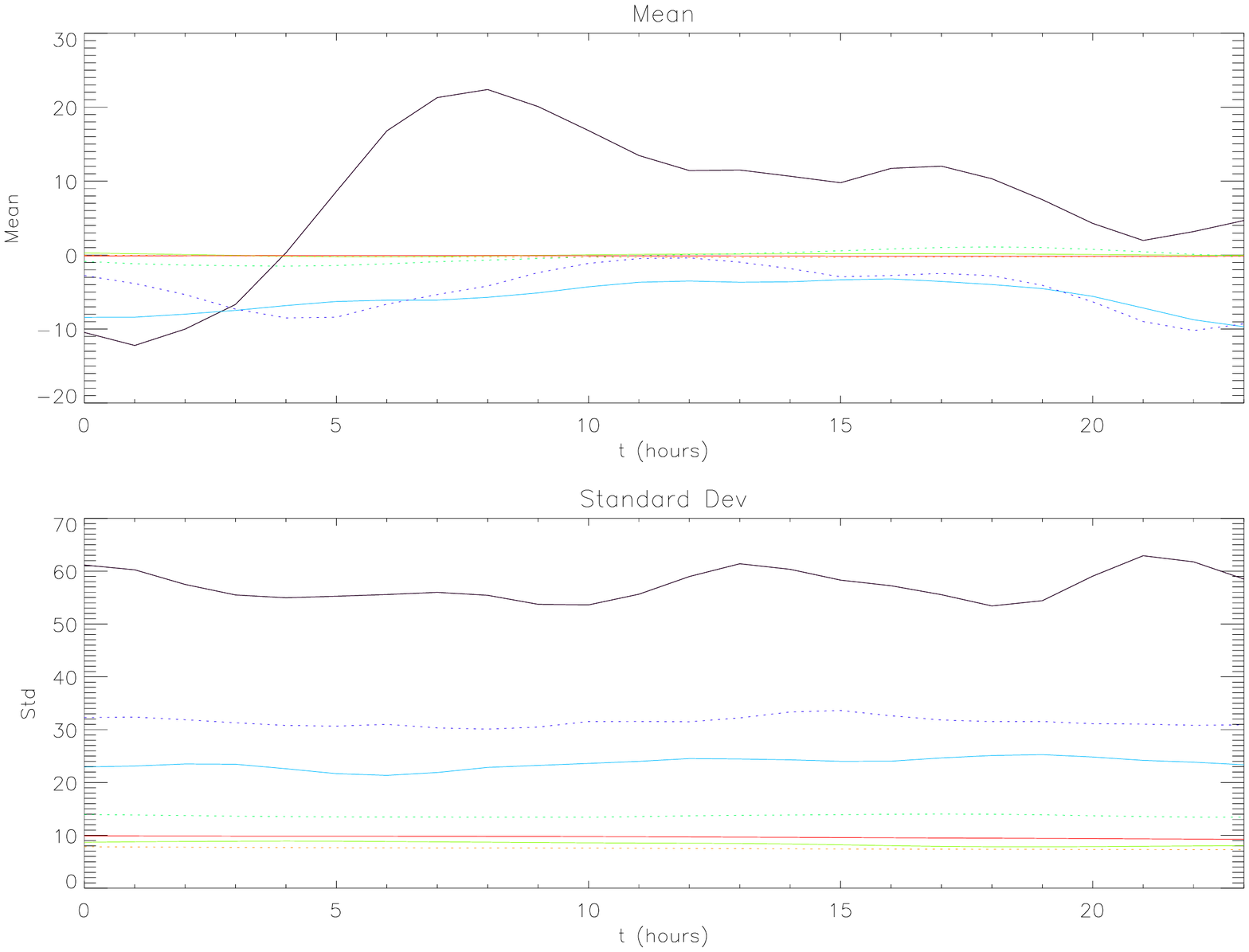}
\includegraphics[angle=0,width=5.4cm,height=7cm]{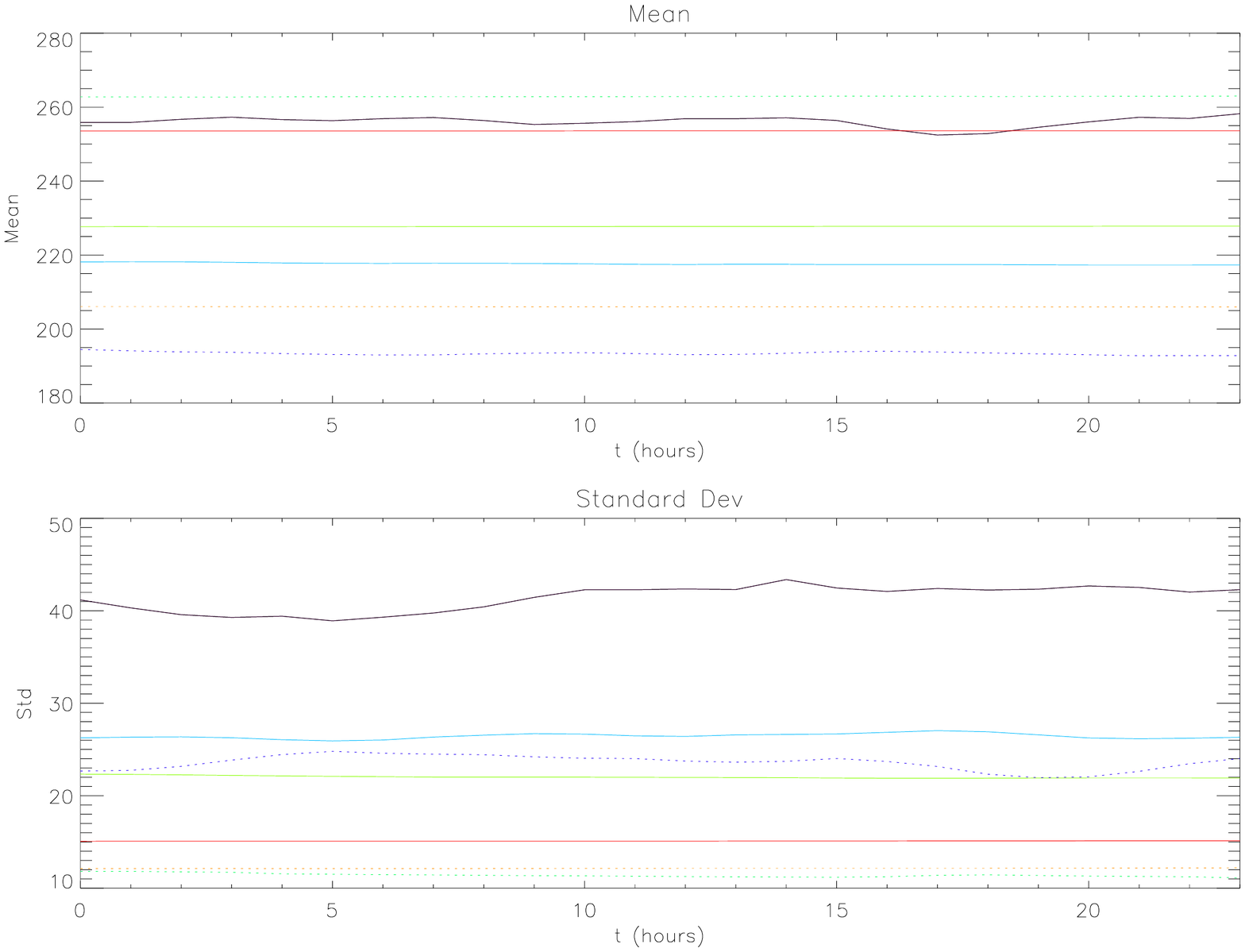}
\caption{Top and bottom panels show the evolution of mean value and standard deviation of zonal flow $U$ (left), meridional flow $V$ (middle) and
temperature $T$ (right). Four solid lines represent four spheres; black, blue, green and red for thermosphere, mesosphere, stratosphere and troposphere, respectively. Three dashed lines represent three pauses; blue, green, and red for mesopause, stratopause and tropopause, respectively.}
\vskip 1cm
\label{Fig4}
\end{figure}

\begin{figure}
%\centering

\includegraphics[angle=90,width=16cm,height=7cm]{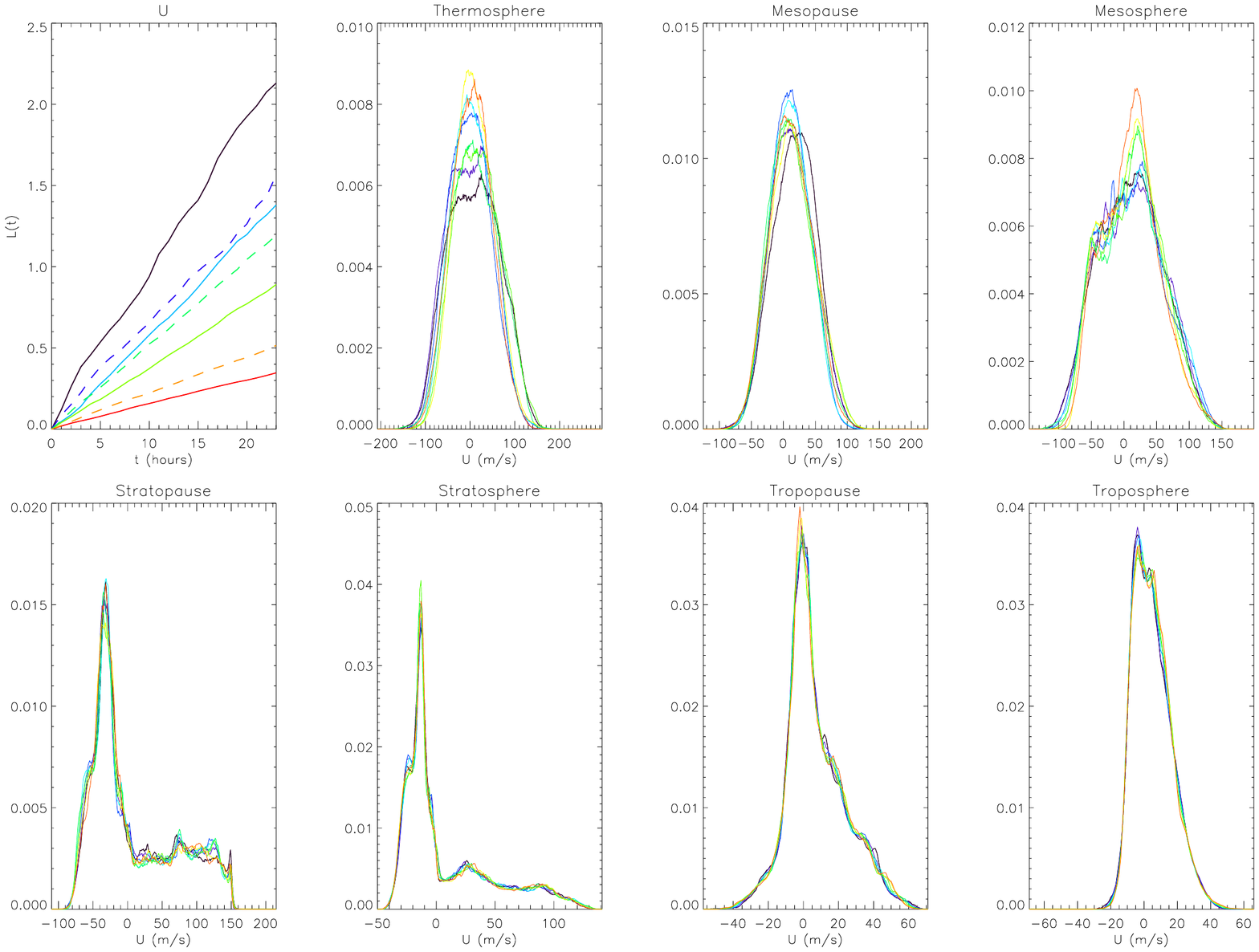}
\includegraphics[angle=90,width=16cm,height=7cm]{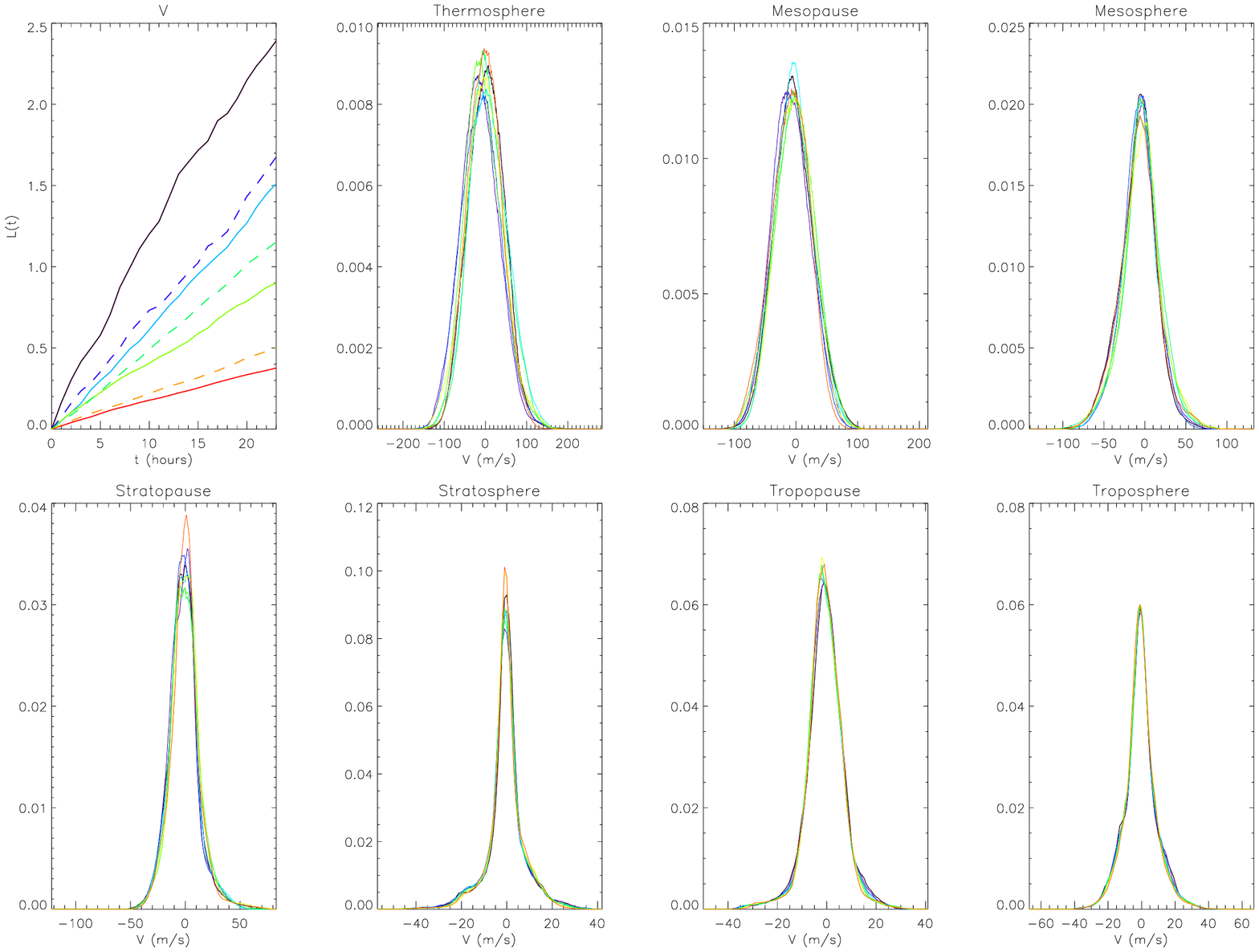}
\includegraphics[angle=90,width=16cm,height=7cm]{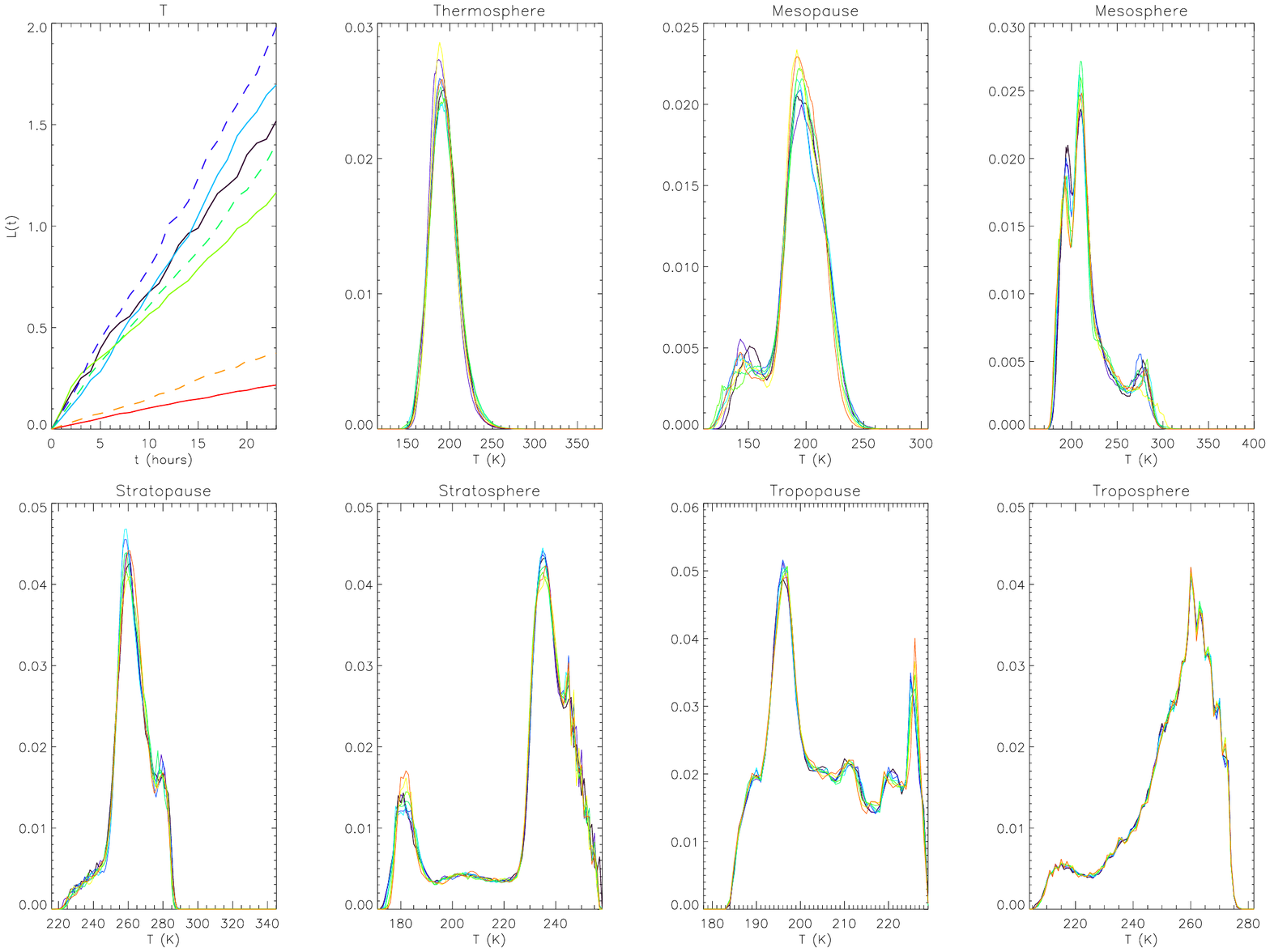}
\caption{Top 8 panels represent ${\cal L}$, time-dependent PDFs at the seven layers for zonal flows $U$;
middle 8 panels are for meridional flow $V$;
bottom 8 panels are for temperature $T$.}
\vskip 1cm
\label{Fig5}
\end{figure}

We use 1-day data at all longitude and latitude, and 5 points in altitude
to construct time PDF around the middle of 7 layers.
From the time-dependent PDFs in each case, we calculate mean value and standard deviation as a function of time.
Fig.~\ref{Fig4} shows the evolution of mean value and standard deviation of zonal flow $U$ (left), meridional flow $V$ (middle) and
temperature $T$ (right). Four solid lines are for the four spheres, black, blue, green and red representing thermosphere, mesosphere, stratosphere and troposphere, respectively. Three dashed lines are for the three pauses, blue, green, and red representing mesopause, stratopause and tropopause, respectively. It is noticeable that in Fig.~\ref{Fig4}, standard deviations of $U$ and $V$ tend to be much larger than mean values at all 7 layers. At any fixed tine, prominent is
a clear phase shift between $U$ and $V$ at the same level. This is due to the presence of strong 
gravity waves, driving an almost isotropic turbulence with the phase shift between $U$ and $V$.  
Also, much less change in the mean temperature compared to its standard deviation is observed. However, there is no systematic variation in either mean or standard deviation from the top to the bottom layers of the atmosphere. This is to be contrasted to the behaviour of the information length, discussed in detail below.\\

The first 8 panels in Fig.~\ref{Fig5} show time-dependent PDFs for zonal flows together with the evolution of the information length in 7 layers. 
Different lines denote PDFs at different times in each panel. From these, it is clear that PDFs are in general non-Gaussian and non-symmetric. 
In comparison, the evolution of information length looks much simpler as ${\cal L}(t)$ against $t$ shows roughly a straight line.
It is interesting that the largest ${\cal L}$ is obtained for the top layer (thermosphere)
and monotonically decreases from the top to the bottom (troposphere). 
\begin{figure}
%\centering
\includegraphics[angle=90,width=16cm,height=7cm]{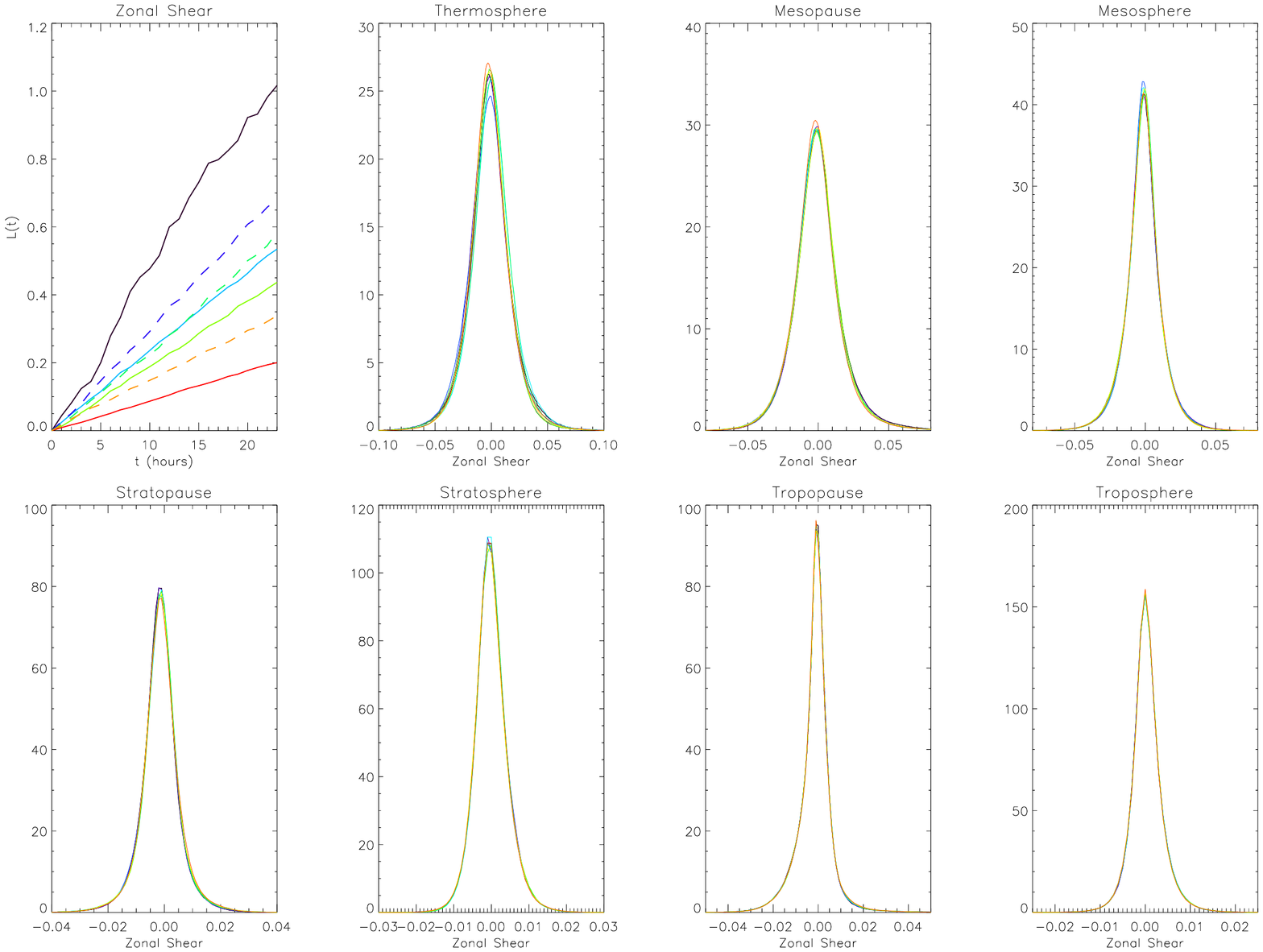}
\includegraphics[angle=90,width=16cm,height=7cm]{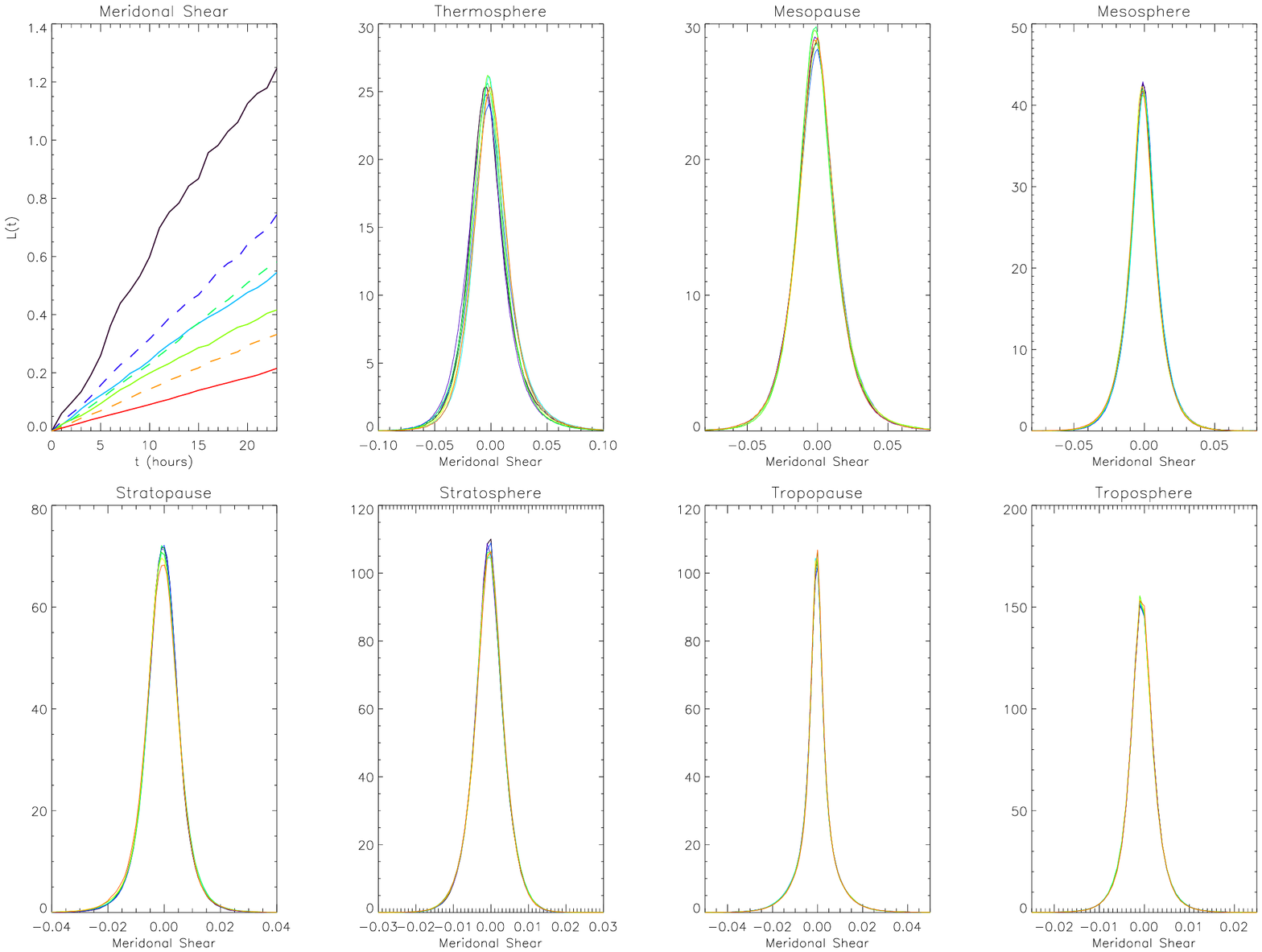}
\includegraphics[angle=90,width=16cm,height=7cm]{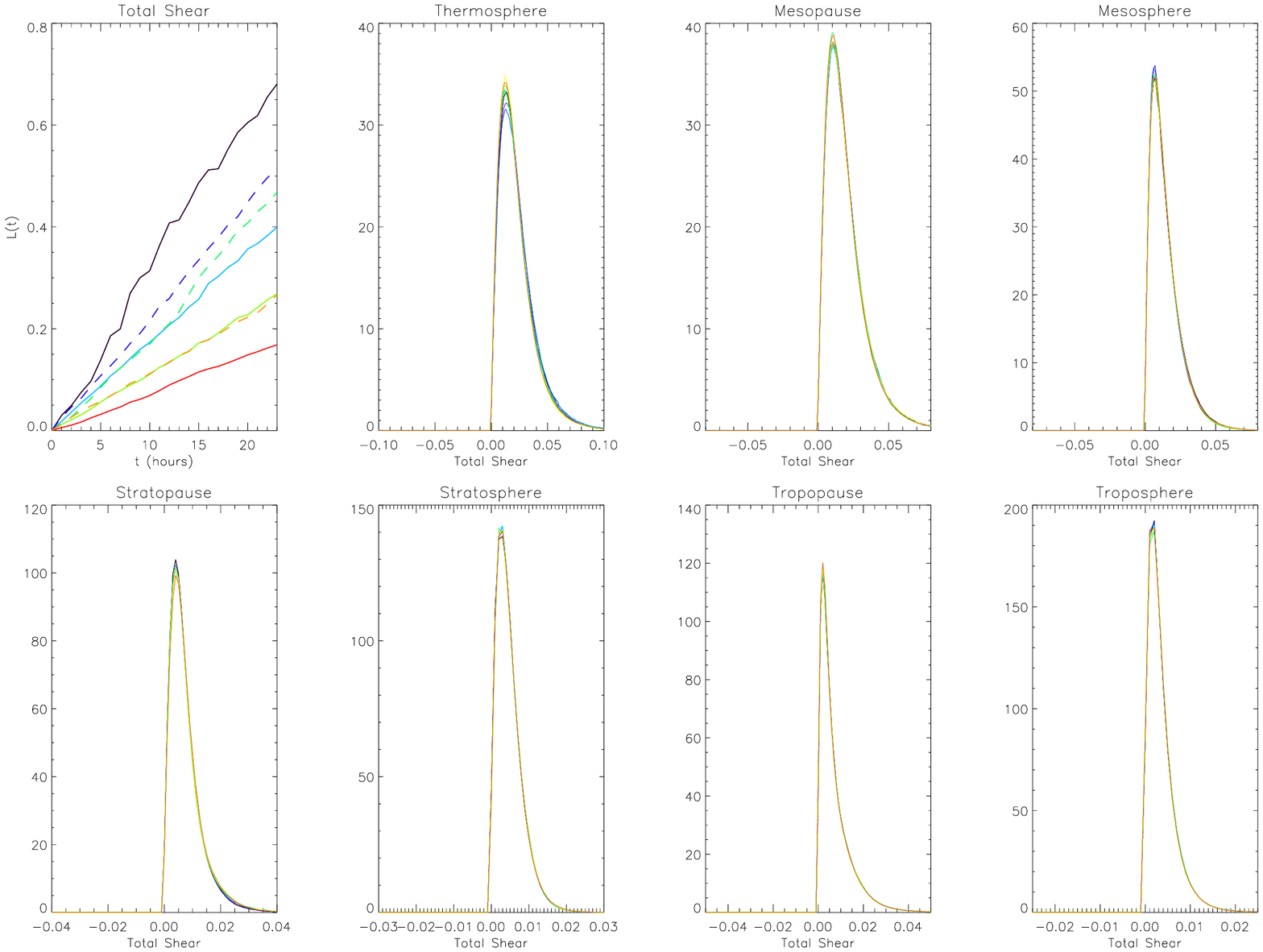}
\caption{Top 8 panels represent ${\cal L}$, time-dependent PDFs at the seven layers for zonal shear;
middle 8 panels are for meridional shear;
bottom 8 panels are for total shear.}
\vskip 1cm
\label{Fig6}
\end{figure}

Equivalent figures are shown for meridional flows $V$ in the next 8 panels in Fig.~\ref{Fig5}. Compared with PDFs of $U$, PDFs of $V$ are much narrower and closer to the Gaussian. Nevertheless, ${\cal L}(t)$ for $V$ behaves similarly to ${\cal L}$ for $U$. In particular,
its value decreases monotonically from the top to the bottom layers of the atmosphere. 
It is also remarkable that the time evolution of ${\cal L}(t)$ is quite similar for $U$ and $V$ at the same layer.
This similarity in ${\cal L}(t)$ results from the similar evolution of the time-dependent PDFs of $U$ and $V$, signifying a strong correlation between $U$ and $V$ due to strong gravity waves (isotropic turbulence), as noted above.

The case of temperature $T$ is shown in the last 8 panels in Fig.~\ref{Fig5}, where we observe quite broad PDFs with more than one peak at some layers.
The ordering of ${\cal L}(t)$ for $T$ is a bit different from that for $U$ and $V$ near the top layer while similar near the bottom layer. 
Also, for $U$ and $V$, the largest information gradient is between thermosphere and mesopause. In comparison, for $T$, the largest information gradient is observed between tropopause and stratopause while thermosphere is well coupled to mesopause with similar ${\cal L}$. \\

In comparison with Fig.~\ref{Fig6}, PDFs of zonal shear, meridional shear and total shear in Fig.~\ref{Fig6} all have much simpler shapes with a narrower width. The evolution and ordering of ${\cal L}$ for all shears is remarkably similar to those for zonal/meridional flows shown in Fig.~\ref{Fig5} although the value of ${\cal L}$ for shears is smaller than for that for flows (due to less change in time-dependent PDFs).
More detailed investigation on the dependence of ${\cal L}$ on the altitude is presented in \S II.3 where ${\cal L}$ is calculated for all pressure level (instead of the 7 levels).

Finally, we check on the robustness of our results by using different 1-day data and also 2-day data for 
zonal flows, meridional flow and temperature and by performing similar analysis.

\subsection{Information budget across latitude}
\begin{figure}
%\centering
\includegraphics[angle=90,width=5.4cm,height=7cm]{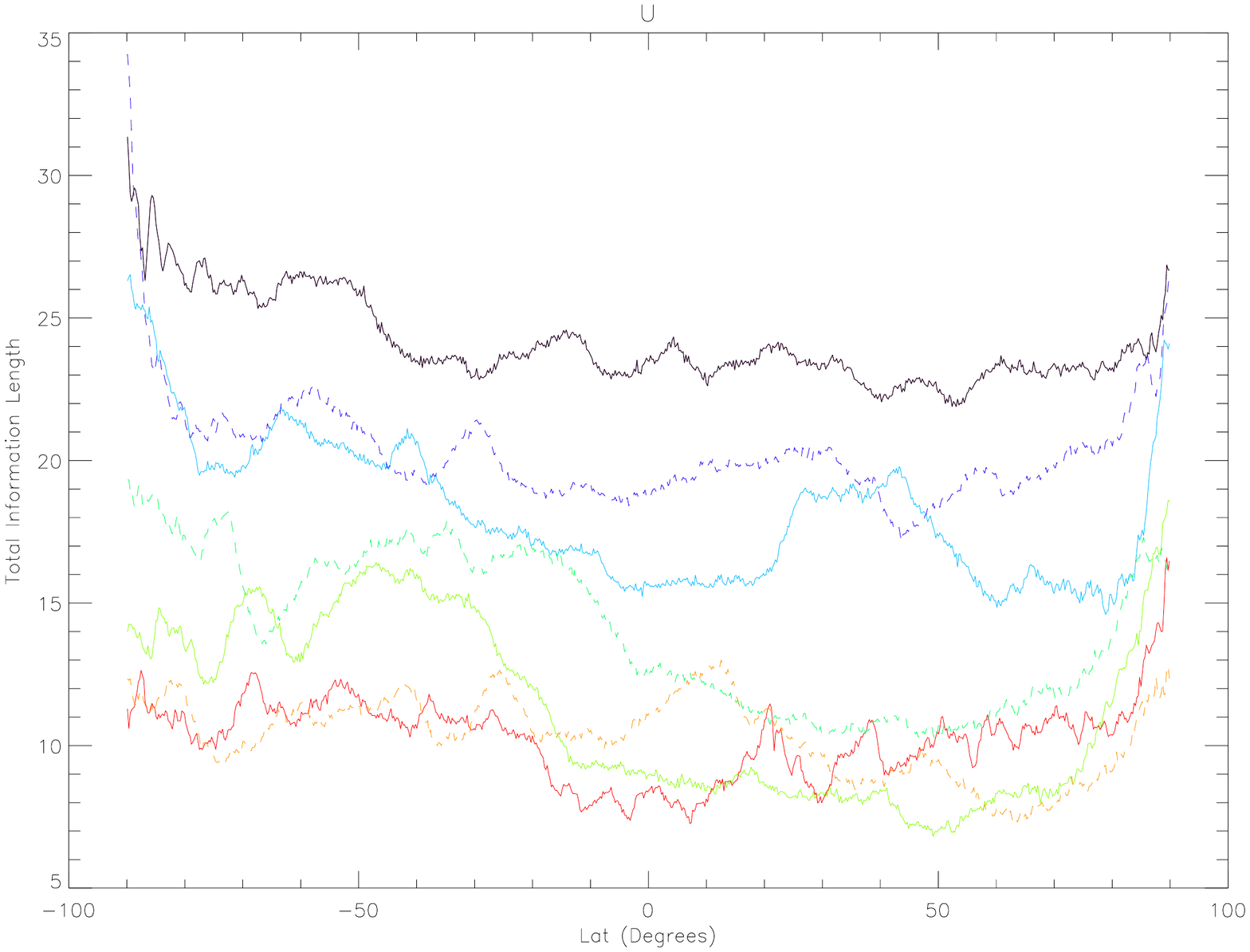}
\includegraphics[angle=90,width=5.4cm,height=7cm]{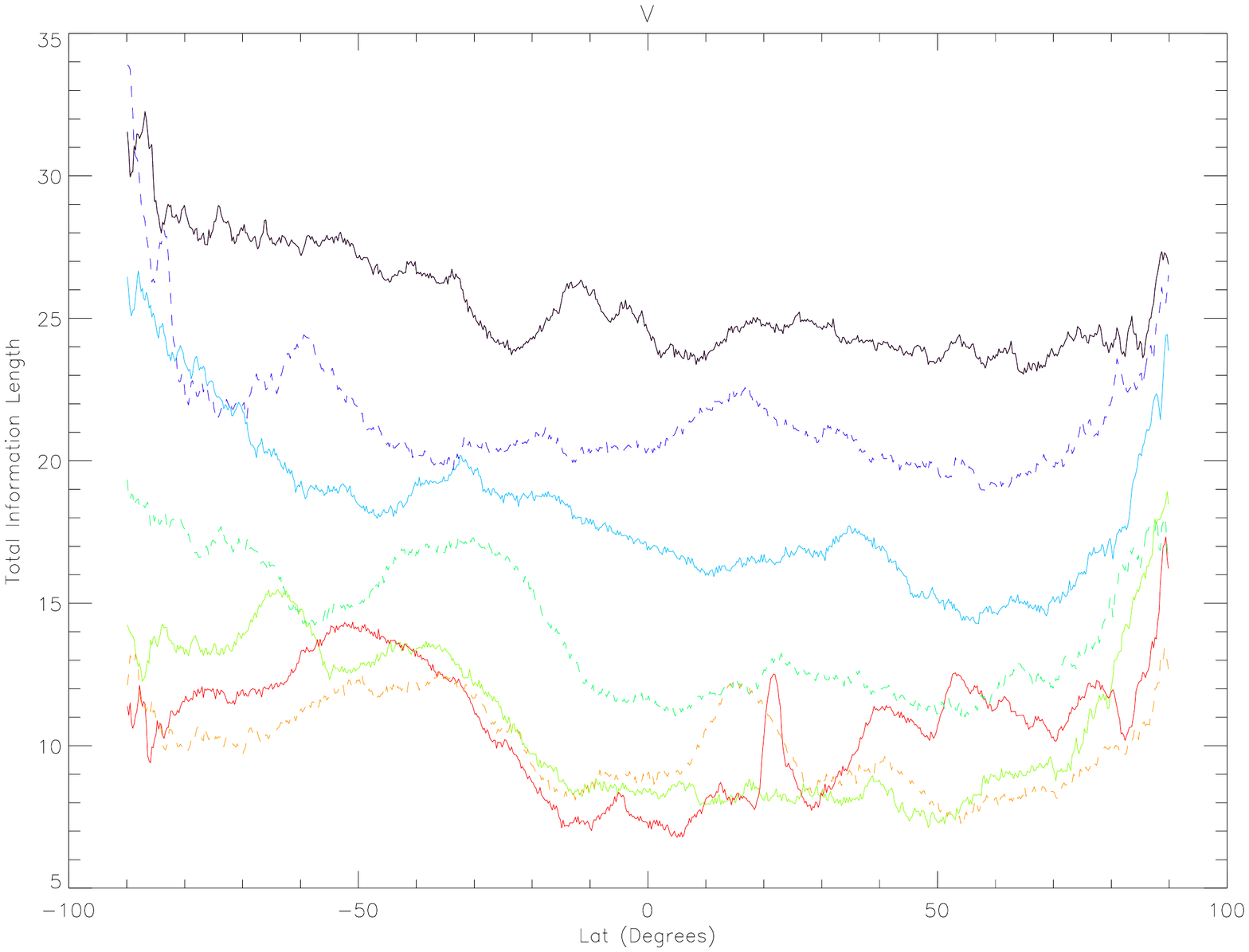}
\includegraphics[angle=90,width=5.4cm,height=7cm]{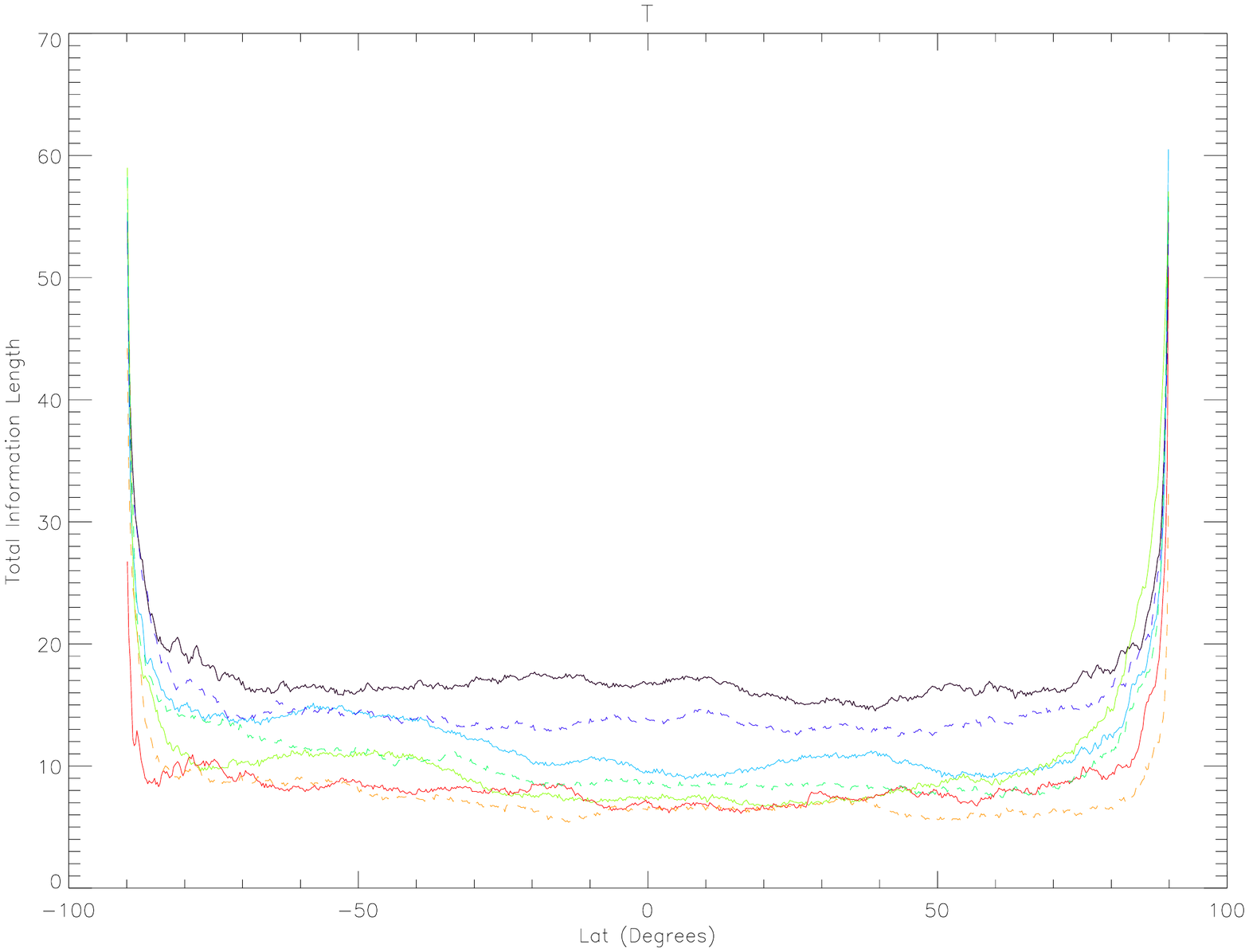}
\includegraphics[angle=90,width=5.4cm,height=7cm]{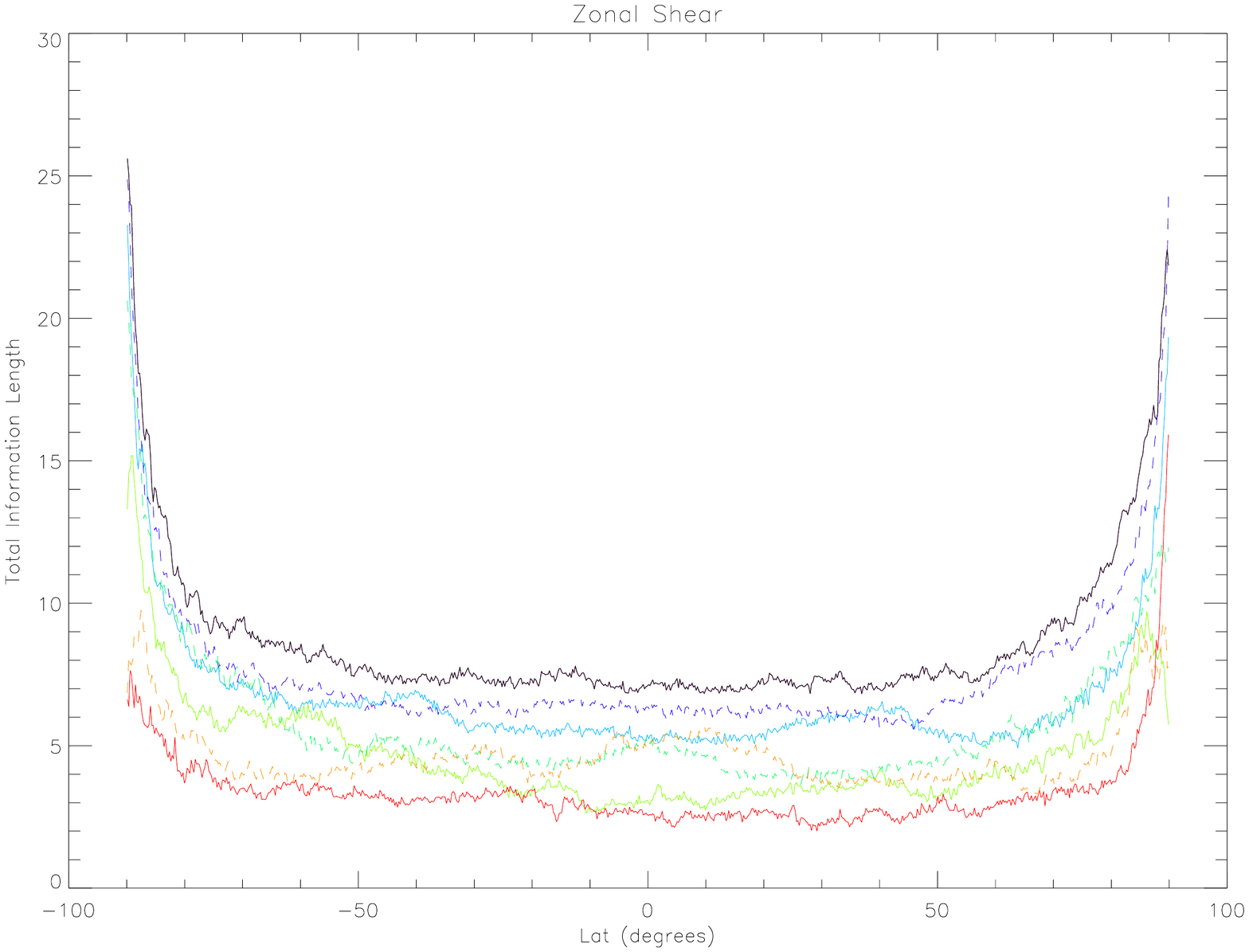}
\includegraphics[angle=90,width=5.4cm,height=7cm]{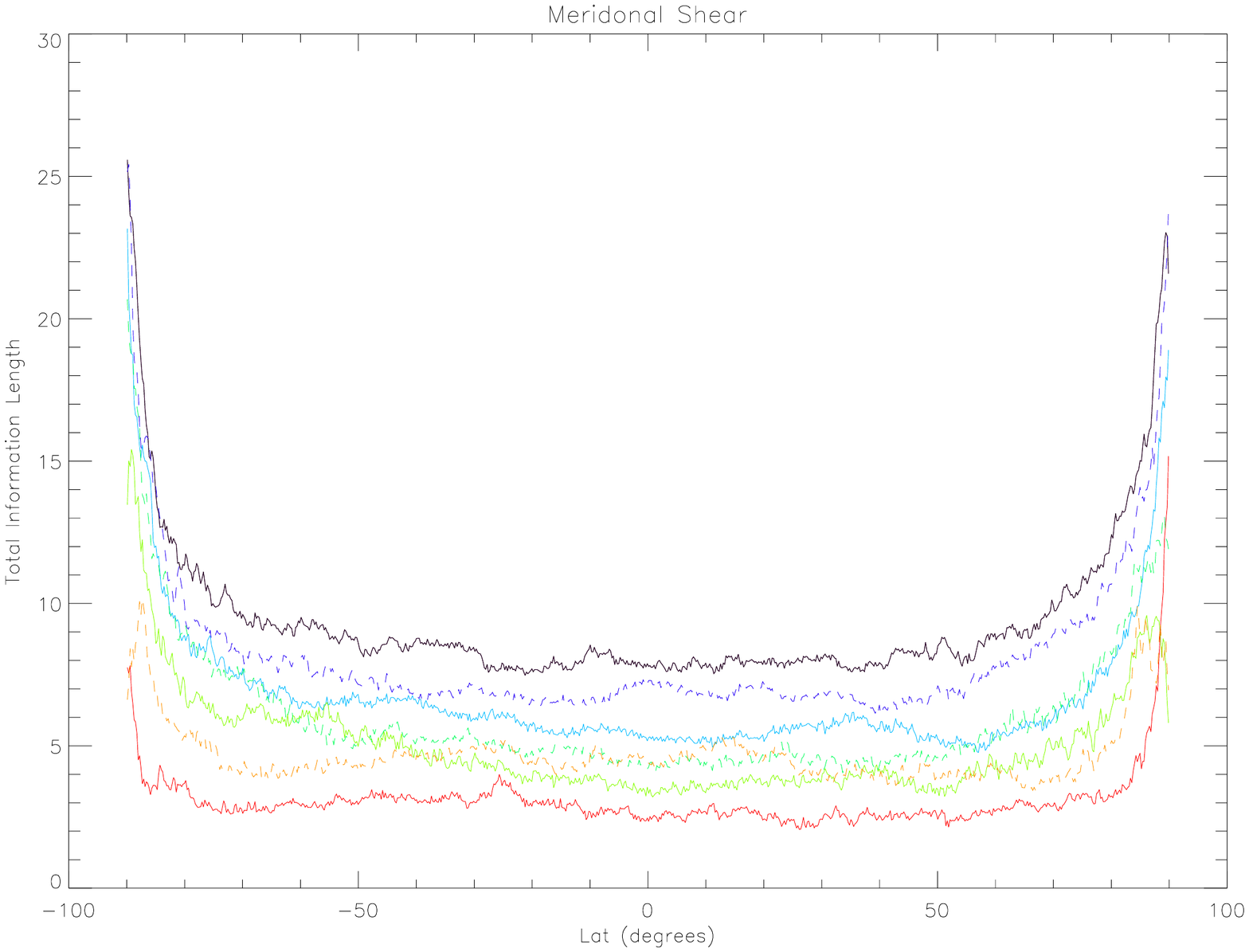}
\includegraphics[angle=90,width=5.4cm,height=7cm]{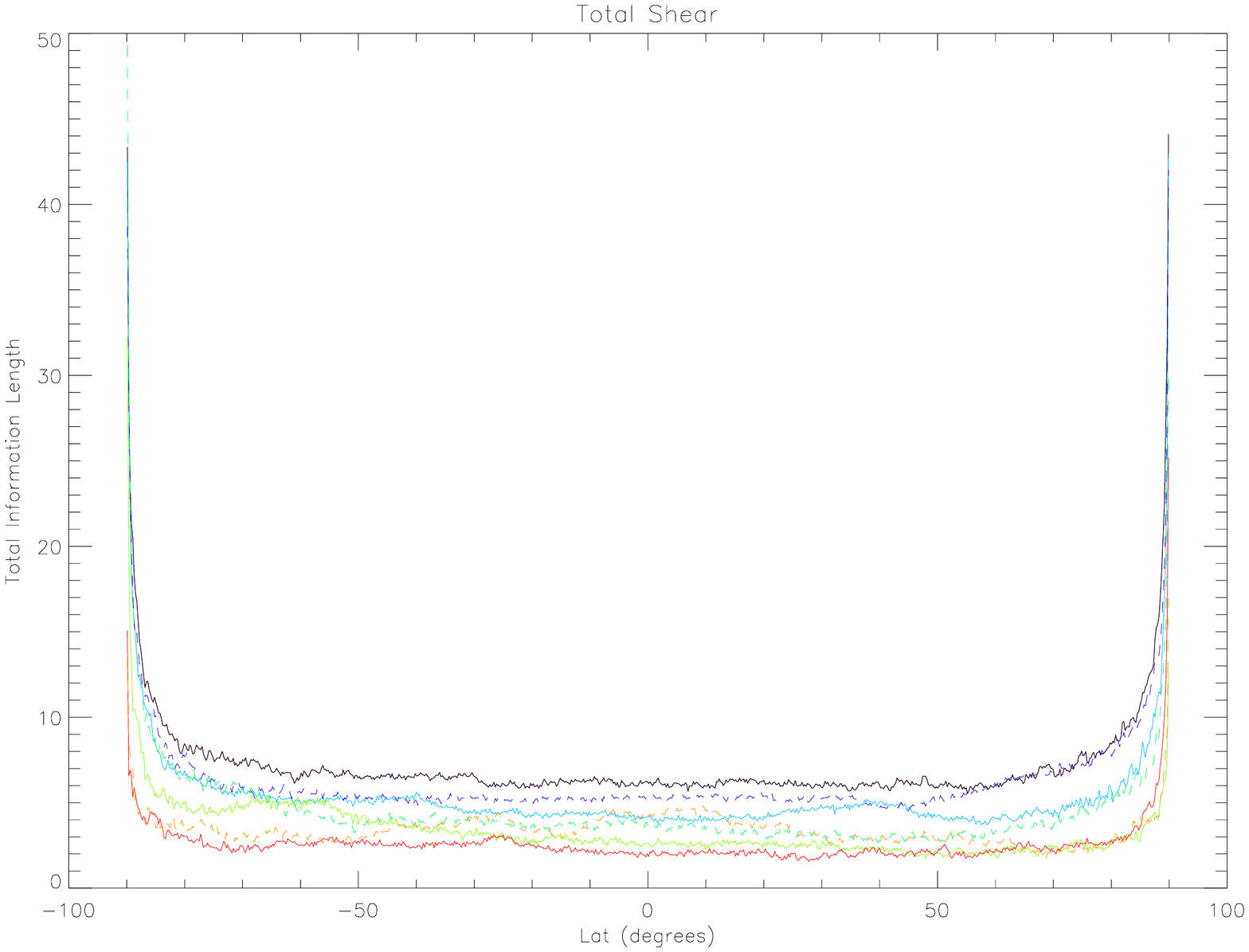}
\caption{Top: the total information length against latitude for U, V, and T from left to right;
Bottom: the total information length against latitude for zonal shear, meridional shear and total shear from left to right.}
\vskip 1cm
\label{Fig7}
\end{figure}

The global information budget studied in \S II.A includes the contribution from all different
latitudes. In order to understand how the information is distributed across latitude,
we now use data from all longitude and 5 points in each layer for each variable.
Fig.~\ref{Fig7} shows the total ${\cal L}$ against latitude, six panels for different
variables. Each panel contains 4 solid lines (for spheres) and 3 dashed lines (for pauses),
using the same colour convention as in Figs.~\ref{Fig5}-\ref{Fig6}.

For all six variables in Fig.~\ref{Fig7},  the total information length reveals an interesting hemispheric asymmetry. Specifically, except for the bottom layer (troposphere in red colour), the information length for $U$, $V$, zonal shear, meridional shear and total shear, tends to be larger in the south than in the north. Exactly the opposite tendency is seen in the information length for $T$. That is, the information length of flows and shears are anti correlated with temperature. Physically, this signifies the role of flows/shear flows in regulating temperature, reminiscent of turbulence regulation by shear flows. At the troposphere, the information length seems to be larger in the north than in the south for all variables, suggesting stronger correlation among them. 

Finally, it is interesting to see that The overall values of ${\cal L}$ in Fig.~\ref{Fig7} are much larger than those Figs.~\ref{Fig5} and \ref{Fig6}. This reflects the fact that the increases in sample size, the information content and thus information length decrease while the uncertainty increases.

\subsection{Information budget across altitude}
\begin{figure}
%\centering
\includegraphics[angle=90,width=4.cm,height=7cm]{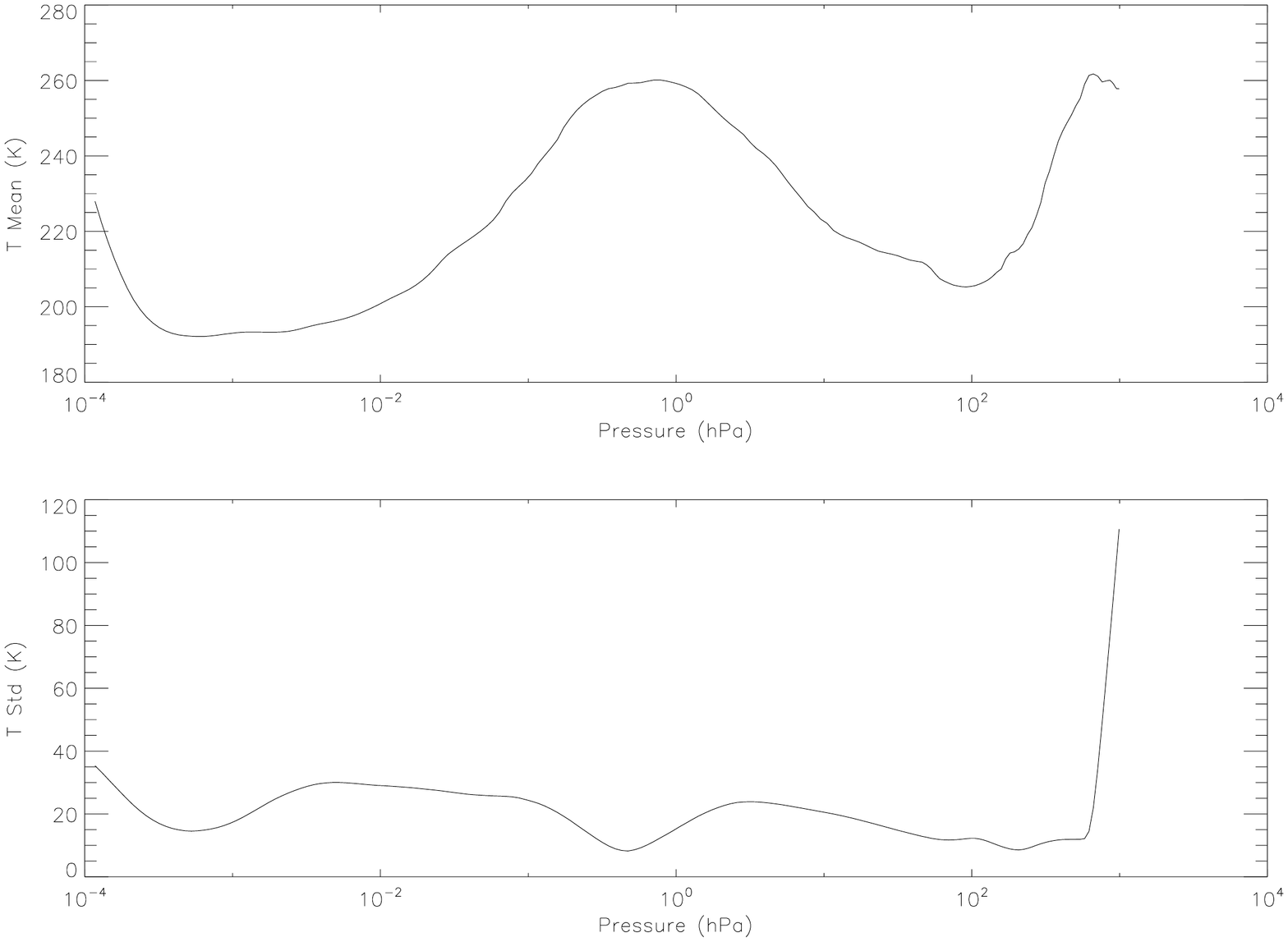}
\includegraphics[angle=90,width=4.cm,height=7cm]{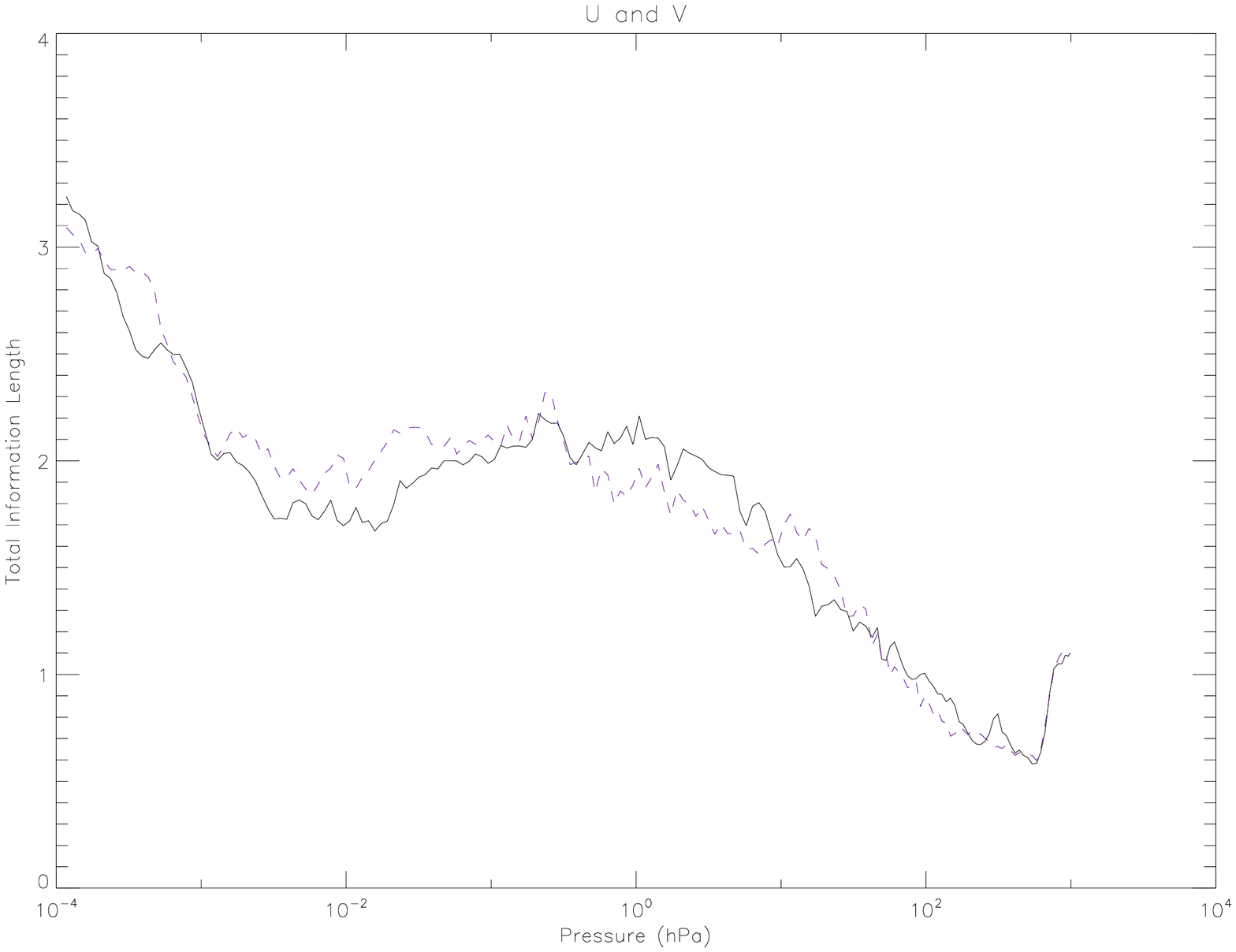}
\includegraphics[angle=90,width=4.cm,height=7cm]{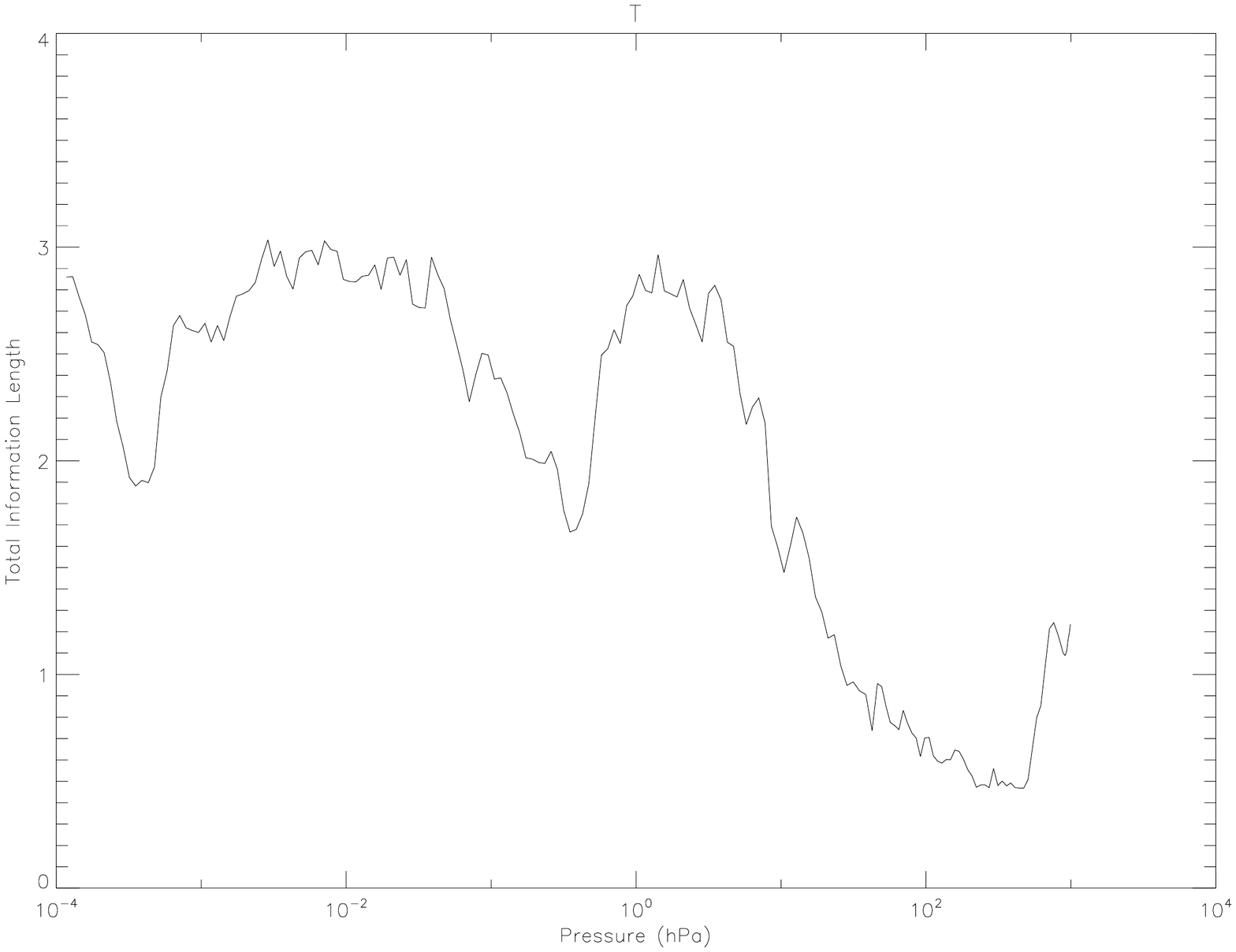}
\includegraphics[angle=90,width=4.cm,height=7cm]{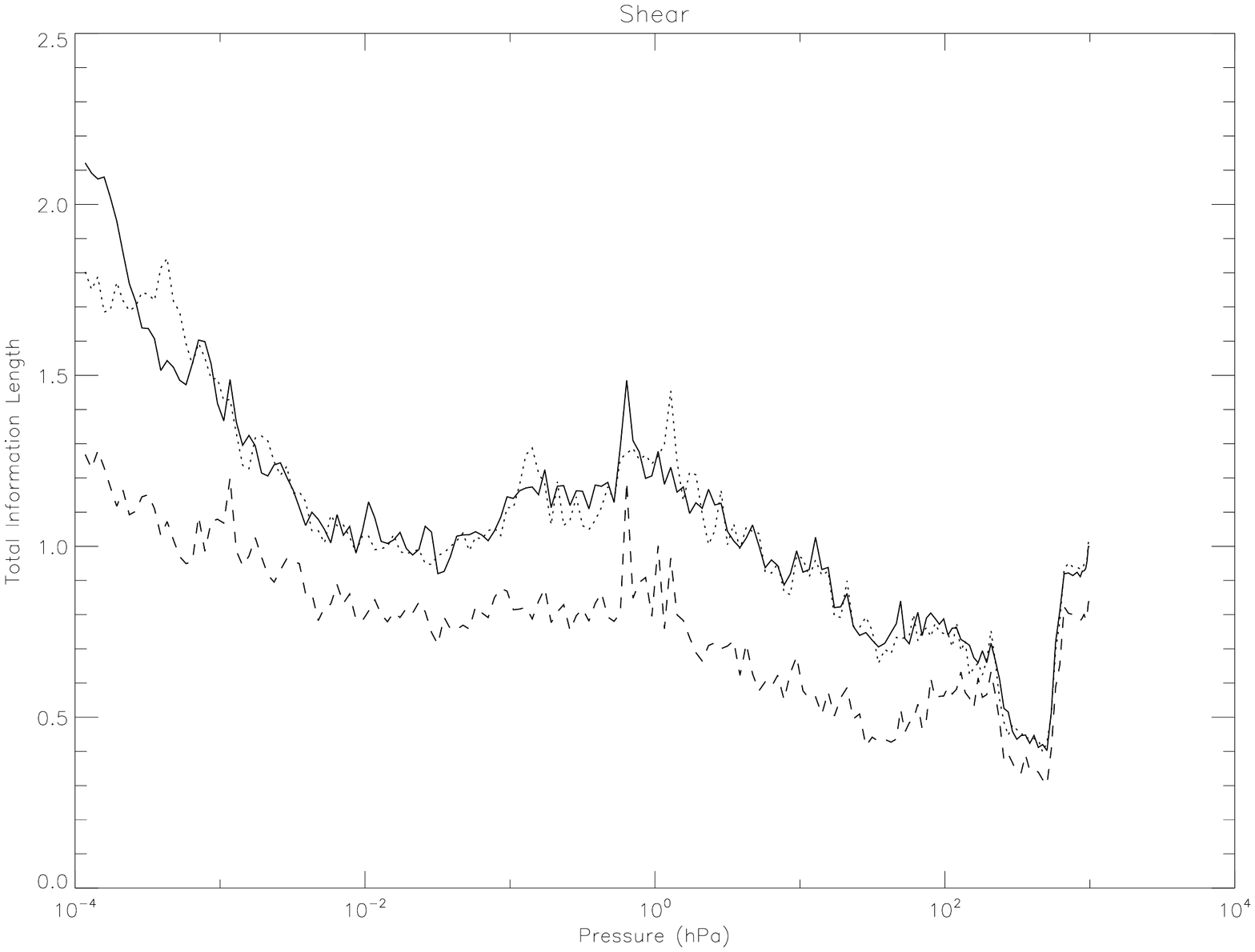}
\caption{First panel: mean temperature (top) and standard deviation (bottom) against pressure; Second, third and last panels are the total information length against pressure for $U$ \& $V$, $T$, and shears (zonal in dotted line, meridional in dashed line, and total shear in solid line)}
\vskip 1cm
\label{Fig8}
\end{figure}

In \S II.A-B, we calculated PDFs using the data at 5 points around the middle of each layer.
To map out the information budget across altitude, we now sample the data at all longitude
and latitude to calculate time-dependent PDFs and ${\cal L}$ at each altitude classified by the pressure level.
Since PDFs now includes only one pressure level, compared to 5 points in \S II.A, the number of data used to calculate PDFs
are 5 times smaller. This will result in smaller value of ${\cal L}$ as discussed later.

The total information length at the end of 1 day against the pressure level is shown in Fig.~\ref{Fig8} for
different variables. Specifically, the second, third and last panels show the total information length against pressure for $U$ \& $V$, $T$, and shears (zonal in dotted line, meridional in dashed line, and total shear in solid line. 
The average mean temperature (top) and standard deviation (bottom) against the pressure level are also shown in the first panel so that we can identify 
physical location of three pauses where the temperature gradient is zero.
It is noteworthy that the standard deviation can significantly vary over the region where
the mean temperature is almost constant. In particular, the standard deviation increases rapidly with the pressure level near the top layer.

The dependence of mean and standard deviation on pressure leads to the  
an interesting non-monotonic behaviour of ${\cal L}$ of $T$ against pressure, shown in the third panel of Fig.~\ref{Fig8}. Specifically, ${\cal L}$ for $T$ shows three distinct local minima, occurring near (but not exactly at) the three pauses. Interestingly, the local minima of ${\cal L}$ of $T$ in fact seems to coincide the local minima of the standard deviation of $T$ in the first/bottom panel. Overall, ${\cal L}$ tends to decreases with the altitude albeit in a complex form due to the presence of the three local minima. This ${\cal L}$ against the pressure gives us a better understanding of ${\cal L}$ against time for the 7 layers in shown in Fig.~\ref{Fig5}. Specifically, the ordering of ${\cal L}$ of $T$ at different layers in Fig.~\ref{Fig5} is the manifestation of
a non-monotonic behaviour of ${\cal L}$ against the pressure level, due to the local minima.

The general tendency of ${\cal L}$ decreasing with the altitude is also observed for $U$, $V$ and shears, local minima occurring at the similar pressure level.
However, unlike ${\cal L}$ of $T$, ${\cal L}$ of flow/shears exhibits only two distinct local minima; as the altitude increases, the first local minimum in ${\cal L}$ for $U$, $V$ and shears occurs between the first two local minima of $T$ while the second local minimum in ${\cal L}$ for $U$, $V$ and shears seems to coincide with the location of the third local minimum of ${\cal L}$ for $T$. This suggests a anti-correlation between flows/shears and $T$ nearer the mesosphere while a stronger correlation nearer the troposphere, similar to the behaviour that we observed in Fig.~\ref{Fig7}.
Furthermore, it is interesting that the two local minima of flows/shears do not occur near the three pauses such a way that ${\cal L}$ for flows/shear at the 7 layers in Figs.~\ref{Fig5} and \ref{Fig6} yet show the monotonic increase of ${\cal L}$ with the altitude.
The overall values of ${\cal L}$ in Fig.~\ref{Fig8} are larger than the corresponding values in Figs.~\ref{Fig5} and \ref{Fig6} at the 7 layers, and this is due to the fact that the information length decreases with the sample size. That is, with the increases in sample size, the information decreases while the uncertainty increases, as noted in \S II.B.

\section{Conclusion and Discussion}
We investigated time-dependent PDFs and information length from the Earth atmosphere data (the global circulation model (WACCM)).
Time-dependent PDFs were in general non-Gaussian, and the information length calculated from these PDFs shed us a new 
perspective of understanding variabilities, correlation among different variables and regions. 
Specifically, we analyzed the WACCM model data and calculated time-dependent PDFs as a function of the altitude (pressure level) and the latitude to investiagate the information budget across different altitudes and latitudes. The information length was found to increase with the altitude albeit in a complex form.
Specifically, the information length for temperature reveals the three local minima coinciding with the location of the local minima of the standard deviation of temperature, near the three pauses. In comparison, the information length for zonal flows, meridional flows, zonal shears, zonal shears, and total shears exhibits only two minima. Much similarity in the behaviour of the information length among flows and shears in the information length, in comparison with the information length for the temperature, suggests a much strong correlation among flows/shears. This means a stronger correlation among flows/shears because of a strong coupling between zonal and meridional flows through gravity waves in this WACCM model.

The correlation between flows/shears and temperature was shown to depend on the altitude, with 
the anti-correlation (correlation) between flows/shears and temperature at a higher (low) altitude. Overall, the information length was found to tend to increases as the latitude increase, with the interesting hemispheric asymmetry for flows/shears/temperature. We propose that the information would tend to flow from a higher information length to a small information length since a small information is due to a large
entropy and since the direction of time follows the direction of the entropy increase. The information flow from higher to lower altitude/latitude in general then highlights the importance of the high latitude and altitude in the information budget in the Earth Atmopshere.  

In summery, our results suggest the utility of the information length as a useful index to understand correlation among different variables and regions as well as information flow.

\section{Acknowledgements}
EK acknowledges the Leverhulme Trust Research Fellowship (RF-2018-142-9); EK and JH acknowledge the HAO visitor programmes for their support and are grateful for the hospitality during their one month visit to HAO.

\appendix

\end{document}